\documentclass[twocolumn, prl, aps, superscriptaddress, longbibliography,showpacs,amsmath,amssymb,floatfix]{revtex4-1}
\usepackage{changes}
\usepackage{graphicx}
\usepackage{dcolumn}
\usepackage{bm}
\usepackage{graphicx}                                   
\usepackage{amssymb}

\usepackage{amsmath}
\usepackage{epsfig}
\usepackage{xcolor}
\usepackage{tabu}
\usepackage{mathtools}
\usepackage[colorlinks,linkcolor=blue,anchorcolor=blue,citecolor=blue,urlcolor=blue]{hyperref}
\usepackage{physics}
\usepackage{float}
\usepackage{diagbox}
\usepackage{inputenc}

\begin{document}

\title{Proof of the exact diffusion constant via first passage time in quasi-periodic potentials}
\author{Ming Gong}
\affiliation{Key Laboratory of Quantum Information, University of Science and Technology of China, Hefei 230026, China}
\affiliation{Anhui Province Key Laboratory of Quantum Network,
University of Science and Technology of China, Hefei 230026, China}
\affiliation{Hefei National Laboratory, University of Science and Technology of China, Hefei 230088, China}
\affiliation{Synergetic Innovation Center of Quantum Information and Quantum Physics, University of Science and Technology of China, Hefei 230026, China}
\date{\today }
\date{\today }
	
\begin{abstract}
Brownian motion in terms of Lifson and Jackson (LJ) formula has been widely explored in periodic systems and it has been believed for a long time that the LJ formula only applies to periodic potentials. Recently we show that for the following Brownian motion $\gamma \dot{x} = -U'(x) + \xi$, where $U(x)$ is the quasi-periodic potential, the effective diffusion constant can still be described by the LJ formula $D^* = D/(\langle \exp(\beta U)\rangle \langle \exp(-\beta U)\rangle)$, where the average is redefined as $\langle \exp(\beta U)\rangle = \lim_{L\rightarrow \infty} L^{-1} \int_0^L \exp(\beta U(x))dx$. In this manuscript we prove this result exactly using the mean first passage time $\tau(x)$, with boundary conditions $\tau(\pm L) = 0$, and show that the effective diffusion constant can be determined using $D^* =\lim_{L \rightarrow \infty} L^2/(2\tau(0))$, where $\pm L$ is the two positions of the absorbing boundary.  We exactly solve the equation of motion of  $\tau(x)$ and obtain the above result with the aid of Jacobi-Anger expansion method. Our result can be generalized to the other potentials and even higher dimensions, which can greatly broaden our understanding of Brownian motion in more general circumstances. The requirement for a well-defined effective diffusion constant $D^*$ in more general potentials is also discussed. 
\end{abstract}
\maketitle

Brownian motion, studying of the erratic random motion of microscopic particles suspended in a fluid, has been one of the most revolutionary concepts in statistical physics \cite{Bian2016111Years, ChandrasekharReview, mel1991kramers, reimann2002brownian, 
hanggi2009artificial, hanggi1990reaction}. In free space, this motion has been well understood since the pioneer works by  Einstein, von Smoluchowski and Langevin in the beginning of the 20th century \cite{einstein1905molekularkinetischen, Bian2016111Years}. In a periodic potential, say $U(x) = U(x+a)$, with $a$ being the period, the Brownian motion described by the Langevin equation can be written as  
\begin{equation}
\gamma \dot{x} = -U'(x) + \xi,
\end{equation}
where $\gamma$ is the viscosity coefficient and the stochastic force satisfies  $\langle \xi(t) \rangle  = 0$,  $\langle \xi(t) \xi(t') \rangle = 2\gamma k_B T \delta(t-t')$. The effective diffusion constant $D^*$  is given by the well-known Lifson and Jackson (LJ) formula \cite{lifson1962self, FESTA1978229,  gunther1979mobility, weaver1979effective}
\begin{equation}
	D^* = \lim_{t\rightarrow \infty} { \langle x^2 \rangle \over 2t} = {D \over \langle \exp(\Phi(x))\rangle\langle \exp(-\Phi(x))\rangle},
	\label{eq-LJ-formula}
\end{equation}
where $\Phi(x) = \beta U(x)$,  $\beta =1/k_B T$ with $k_B$ and $T$ being the Boltzmann constant and temperature, respectively, and $\langle x^2\rangle$ is the variance of position during random motion. Here the average is defined in a full period as  \cite{weaver1979effective}
\begin{equation}
\langle \exp(\pm \Phi(x)) \rangle = {1\over a} \int_0^a\exp(\pm\Phi(x)) dx.
\label{eq-average0}
\end{equation}
It has been widely believed that the periodicity is necessary for the above definition of $D^*$. In the rough potential, it has been argued by Zwanzig   \cite{Zwanzig1988Diffusion} that when the system has two different periods, say $a$ and $b$, then the above definition may become valid (to some extent of approximation) when and only when one of the period is much larger than the other period, say $a \gg b$. 

Recently  in Ref. \cite{SangYang2025} we show that the above definition can be extended to the more general potentials. We consider the quasi-periodic potential and  show that the above LJ formula can still be applied to determine the effective diffusion constant,  yet the average should be taken in a sufficiently large interval, that is  
\begin{equation}
\langle \exp(\pm \Phi(x)) \rangle = \lim_{L \rightarrow \infty} {1\over L} \int_0^L\exp(\pm \Phi(x)) dx,
\label{eq-average1}
\end{equation}
assuming both limit exist.  This average naturally includes the result of Eq. \ref{eq-average0}.  We demonstrate this result using several different approaches, in which our key argument  is that since the quasi-periodic potential $U(x)$ can be approximated using periodic potential with arbitrary accuracy  with a sufficient large period $L$, then we can still use the LJ formula to derive $D^*$. Furthermore, we show that if this diffusion constant is well defined, then the whole probability function $p(x, t)$ should be related to the Gaussian function in some way   \cite{SangYang2025, sivan2018probability, defaveri2023brownian} 
\begin{equation}
p(x, t) = \mathcal{A} \exp(-\Phi(x)) {1\over \sqrt{4\pi D^* t}} \exp(-{x^2 \over 4D^* t}), 
\label{eq-Pxt}
\end{equation}
in which $\mathcal{A}$ is determined by the normalization condition.  The accuracy of these two parts have been numerically verified.  With this solution, the effective diffusion constant can be obtained from the  Fokker-Planck equation. 

To make our conclusion much more rigorous, it is necessary to prove the above conclusion in terms of mean first passage time (MFPT) $\tau(0)$ starting at $x_0 = 0$ with two absorbing boundaries at $\pm L$, in which the diffusion constant $D^*$ can be defined using 
\begin{equation}
D^* = \lim_{L \rightarrow \infty} {L^2 \over 2\tau(0)}.
\label{eq-Deff}
\end{equation}
This is one of the most standard approach to derive the LJ formula, yet in periodic system $L$ is set to the period $a$.  This new definition will make the boundary condition to be insignificant anymore. Here we would like to thank the referee during the review of Ref. \cite{SangYang2025}, in which he/she insisted that our proof is not exact and the periodicity of the potential is required or $a \gg b$ should be satisfied in using of the LJ formula \cite{Zwanzig1988Diffusion}, thus the effective $D^*$ should be proven using MFPT, leading to this manuscript. 


Let us assume the probability to be $p(x, t | x_0)$, where $x_0$ is the initial position with two absorbing boundaries at $\pm L$. This corresponds  diffusion equation  reads as \cite{Gitterman2000Mean, Dybiec2006Levy}
\begin{equation}
{\partial p(x, t|x_0) \over \partial t} = D{\partial^2  p(x, t|x_0) \over \partial x^2},
\end{equation}
subject to the following boundary conditions 
\begin{eqnarray}
p(x, 0|x_0) = \delta(x-x_0),  \quad 
p(\pm L, t|x_0) = 0.
\end{eqnarray}
We assume $D$ to be position independent, yet its generalization to position dependent is straightforward.  We have to solve the partial differential equation with proper boundary conditions,  which makes $p(x, t|x_0)$ totally different from that in Eq. \ref{eq-Pxt}.  With this we can define the survival probability in the interval $[-L, L]$ as 
\begin{equation}
s(t, x_0) = \int_{-L}^{L} p(x, t|x_0)dx.
\label{eq-stx0}
\end{equation}
Obviously, $s(0, x_0) = 1$, and when $t$ approaches infinity, the particle should inevitably hit the boundaries, thus $s(\infty, x_0) = 0$. Physically, it should be a monotonic decreasing function of $t$, thus we can define $\rho = \rho(t, x_0) = -\partial s/\partial t$ as a probability distribution function, that is  $\int_0^\infty \rho(t, x_0) dt = 1$ for all $x_0 \in (-L, L) $, with $\rho(t, x_0) \ge 0$.  In this way we can calculate the MFPT as \cite{weaver1979effective}
\begin{equation}
\tau(x_0) =  \langle t\rangle = \int_0^\infty t \rho(t, x_0) dt.
\label{eq-taux0}
\end{equation}
We have $a^2 = 2D^*\tau(0)$ in periodic potentials via Eq. \ref{eq-Deff}.  

We can calculate the solution of $p(x, t|x_0)$ directly. The general solution with the absorbing boundary conditions, in terms of plane wave basis, can be written as
\begin{equation}
p(x, t|x_0) = \sum_{n=1}^\infty c_n(t) \sin({n \pi(x-L) \over 2L}),
\end{equation}
with $\dot{c}_n(t) = -n^2 \pi^2 D/(4L^2) c_n(t)$.  With the initial  condition that $p(x, 0|x_0 = \delta(x-x_0))$, we  obtain 
\begin{eqnarray}
p(x,t|x_0)  && =  {1\over L}  \sum_{n=1}^\infty  \exp(-{n^2 \pi^2 D t \over 4L^2})  \nonumber \\ 
&& \sin({n \pi(x_0-L) \over 2L}) \sin({n \pi(x-L) \over 2L}).
\end{eqnarray}
A direct calculation shows that $\int_{-L}^L p(x, t|x_0) \ne 1$ when $t\ne 0$ due to the escape of particle from this finite interval.  Then we can calculate the survival probability  as 
\begin{equation}
s(t, x_0) = \sum_{n \in 2\mathbb{Z}+1} -{4 \over n\pi} \exp(-{n^2 \pi^2 D t \over 4L^2}) \sin({n \pi(x_0-L) \over 2L}),
\end{equation}
where $2\mathbb{Z} +1$ means that only the odd terms (1, 3, 5, $\cdots$) contribute to the total survival probability.  Obviously, when $t$ is large enough, we should have 
\begin{equation}
s(t, 0) \rightarrow {4 \over \pi} \exp(- {\pi^2 D \over 4L^2} t). 
\label{eq-St0}
\end{equation}
With this solution we can determine the probability distribution of first passage time using $\rho = -\partial s(t, x_0)/\partial t$, yielding
\begin{equation}
\tau(x_0)  = \sum_{n \in 2\mathbb{Z}+1} -{4 \over n\pi} {n^2 \pi^2 D  \over 4L^2}  {(4L^2)^2 \over n^4 \pi^4 D^2}  \sin({n \pi(x_0-L) \over 2L}),
\end{equation}
which obviously satisfy the boundary conditions $\tau(x_0 = \pm L) = 0$. The above expression yields the MFPT 
\begin{equation}
\tau(x_0) = {L^2 - x_0^2 \over 2D},
\label{eq-taux0}
\end{equation}
with the aid of the following identity
\begin{equation}
\sum_{n = 1, 3, 5, \cdots} -{4 \over n\pi} {4 \over n^2 \pi^2} \sin(nA) = -{1\over 2} + {2\over \pi^2} (A - {\pi \over 2})^2.
\end{equation}

Hereafter,  let us change $x_0$ to $x$. It has been prove that the MFPT satisfies the following differential equation \cite{weaver1979effective}
\begin{equation}
D  e^{-\Phi(x)} {\partial \over \partial x}  e^{\Phi(x)}   {\partial \over \partial x}  \tau(x) = -1, 
\label{eqDphix}
\end{equation}
with boundary conditions $\tau(x = \pm L) = 0$. This equation is valid for the first-order Langevin equation with time independent potential, without the requirement that $U(x)$ should be a periodic functions.   Physically, $\tau(x)$ is the averaged first passage time the system will hit the boundaries $\pm L$. This definition is different from the diffusion constant defined in the Brownian motion, in which the time is fixed, and the variance of position $\langle x^2\rangle$ is calculated.  

(I) In free space we have $U(x) = 0$,  the above equation for MFPT is reduced to 
\begin{equation}
D {\partial^2 \over \partial x^2} \tau(x) = -1,  \quad \tau(x) = {L^2 - x^2 \over 2D}, 
\label{eq-DUx0}
\end{equation}
which has been obtained in Eq. \ref{eq-taux0}.  We therefore have Eq. \ref{eq-Deff}.  


(II) With the above result, we then consider Brownian motion in the following periodic potential
\begin{equation}
\Phi(x) = U \cos(x),
\end{equation}
with period $a = 2\pi$. We can move $e^{-\Phi(x)}$ to the right hand side and make an integration in $[0, x]$, yielding 
\begin{equation}
D( e^{\Phi(x)}   {\partial \over \partial x}  \tau(x) - B_0) = - \int_0^x e^{\Phi(x)}  dx,
\end{equation}
where $B_0 = e^{\Phi(x)}   {\partial \over \partial x}  \tau(x)|_{x= 0}$.  Use the Jacobi-Anger expansion  $\exp(U \cos(x)) = \sum_n I_n(U) \exp(inx)$,  we  have 
\begin{equation} 
D(e^{\Phi(x)}   {\partial \over \partial x}  \tau(x) - B_0) = - I_0(U)x - \sum_{n \ne 0} I_n(U) g_n(x),
\end{equation}
where $g_n(x) =  (e^{inx} -1)/(in)$.  Obviously, $g_0(x) = x$ and when $n \ne 0$,  $|g_n(x)| \le {2 \over |n|}$. It should be noticed that the modified Bessel function $I_n(U)$ will quickly decays to zero with the increasing of $n$. The second term, which is an oscillation function of $x$, can be estimated using  
\begin{equation} 
 |\sum_{n \ne 0} I_n(U) {e^{inx} -1 \over in}| \le 
\sum_{n \ne 0} {2|I_n(U)| \over n}, 
\end{equation}
which is a finite upper bound independent of $x$. Thus in the large $x$ limit, we have 
\begin{equation}
\int_0^x e^{\Phi(x)}  dx \rightarrow I_0(U)x.
\end{equation} 
This average has a very clear physical meaning, which will be useful for us to understand the results in the quasi-periodic potentials. Physically $I_0(U)$ is nothing but the average of $e^{\Phi(x)} $ in the whole regime, that is 
\begin{equation}
I_0(U) = \langle e^{\Phi(x)}  \rangle = \lim_{x\rightarrow \infty} {1\over x} \int_0^x e^{\Phi(x')}  dx'.
\end{equation}
This function can also be obtained in the periodic system averaged in a full period; however, the above definition has the advantage that the boundary condition is insignificant anymore in the large $x$ limit, which will become essential in the quasi-periodic potentials. 

Now, let us move $e^{\Phi(x)}$ to the right hand side again and we  have 
\begin{eqnarray}
D\int_0^L {\partial \over \partial x}\tau(x)dx  
&=& DB_0 \int_0^L  e^{-\Phi(x)} dx  \\ 
&-& \sum_{n} I_n(U) \int_0^L e^{-\Phi(x)}g_n(x)dx. \nonumber 
\end{eqnarray}
Using the boundary condition that $\tau(L) = 0$, we should have the MFPT with initial point at $x = 0$ as
\begin{equation}
D\int_0^L {\partial \over \partial x}\tau(x)dx = -D\tau(0).
\label{eq-Dtau0}
\end{equation}
Thus using the Jacobi-Anger expansion and we have
\begin{eqnarray}
-D\tau(0) &=& DB_0 (I_0(U) L + \sum_{n \ne 0} I_n(-U) g_n(L) \nonumber \\ 
&-&I_0(U)I_0(U) {L^2 \over 2} -I_0(U) \sum_{n \ne 0} k_n(L),
\\ 
&-& \sum_{n\ne 0,m} I_n(U) I_m(-U) \int_0^L e^{imx} g_n(x)dx, \nonumber
\end{eqnarray} 
where we have defined 
\begin{eqnarray}
k_n(L)  = I_n(-U) \int_{0}^L \exp(inx) x dx. 
\end{eqnarray}
Using $|g_n(L)| \le 2/|n|$, which is a bounded function when $n \ne 0$.  We find that in the large $L$ limit, $k_n(L)  =I_n(-U) (L + {g_n(L) \over (in)^2}) 
\rightarrow I_n(-U)L$, and the next-leading terms will become negligible. 

In the similar way, we can calculate $\int_0^L e^{imx} g_n(x)dx$ exactly, which is an oscillating function of $x$, yielding that this term should at most proportional to $L$. This can be estimated using the following way
\begin{equation}
|\int_0^L e^{imx} g_n(x)dx| \le \int_0^L 
|e^{imx} g_n(x)|dx \le {2\over |n|}L.  
\end{equation}
Collecting all these results together, we find that 
\begin{equation}
-D\tau(0) = -I_0(U) I_0(-U) {L^2 \over 2} + CL + \cdots, 
\end{equation}
where $C$ is related to $DB_0 I_0(U)$ and $-I_0(U) I_n(-U)L$ {\it etc}. Here the suspension points represent the next sub-leading terms with amplitude much smaller than the second term $CL$ in the large $L$ limit. In this way, in the large $L$ limit  $-D\tau(0) \rightarrow -I_0(U) I_0(-U) {L^2 \over 2}$, 
yielding the effective diffusion constant give by the LJ formula in terms of MFPT as 
\begin{equation}
D^* = \lim_{L \rightarrow \infty} {L^2 \over 2\tau(0)} = {D \over \langle \exp(\Phi)\rangle \langle \exp(-\Phi)\rangle}.
\end{equation}
Here the average is taken in a full period $[0, a]$; however when it is averaged in much longer interval $[0, L]$ when $L$ is large enough, we can obtain the same expression.


(III) We next want to derive the effective diffusion constant in some more general quasi-periodic potentials, in which the simplest form can be written as \cite{Lopez2020Enhanced}
\begin{equation}
\Phi(x) = U_a \cos({2\pi x \over a}) + U_b \cos({2\pi x \over b}),
\label{eq-UaUb}
\end{equation}
using the notation in Ref. \cite{SangYang2025}.  It should be noticed that in our model, $a$ and $b$ are not commensurate period, thus $a/b$ is an irrational number. We do not require $a \gg b$ or $b \gg a$.   Using the Jacobi-Anger expansion we will have 
\begin{equation}
\exp(\Phi(x)) = \sum_{n, m} I_n(U_a) I_m(U_b) e^{i({n \over a}  + {m \over b})x}.
\end{equation}
For convenience, let us define ${\bf k} = (n, m)$, ${\bf r} = (1/a, 1/b)$, ${\bf k} \cdot {\bf r} =n/a  + m/b$, and $I_{{\bf k}} = I_n(U_a) I_m(U_b)$, then  $\exp(\Phi(x)) = \sum_{{\bf k}} I_{\bf k} e^{i{\bf k} \cdot {\bf r} x}$.  We write it in this way so that the whole expressions in following will be similar to that with a single period, from which the generalization is rather straightforward. For instance,  we have 
\begin{eqnarray}
D(e^{\Phi} {\partial \over \partial x} \tau(x) - B_0) = - \sum_{\bf k} I_{\bf k} g_{{\bf k}}(x),
\end{eqnarray}
with $g_{\bf k}(x) = (e^{i{\bf k} \cdot {\bf r} x} -1)/(i {\bf k} \cdot {\bf r})$. Similarly when ${\bf k} \ne 0$ 
\begin{eqnarray}
	|\sum_{{\bf k} \ne 0} I_{{\bf k}} {e^{i{\bf k} \cdot {\bf r} x} -1 \over i{\bf k} \cdot {\bf r} }| \le\sum_{n,m \ne 0} {2|I_n(U_a)I_m(U_b)| \over |{\bf k} \cdot {\bf r} |},
\end{eqnarray}
which is finite and independent of $x$, thus we should have 
\begin{equation}
	\int_0^x e^{\Phi(x)}  dx \rightarrow I_0(U_a)I_0(U_b)x.
\end{equation}
Notice that $a$ and $b$ are two incommensurate periods, so $\bf{k}=0$ when and only when $n=m=0$. 

Now, let us move $e^{\Phi(x)}$ to the right hand side again and by using the same procedure we will have
\begin{eqnarray}
	D\int_0^L {\partial \over \partial x}\tau(x)dx  
	&=& DB_0 \int_0^L  e^{-\Phi(x)} dx \nonumber \nonumber\\ 
	&-& I_0(U_a)I_0(U_b) \int_0^L e^{-\Phi(x)} xdx \nonumber\\ 
	&-& \sum_{\bf k\neq 0} I_{\bf k} \int_0^L e^{-\Phi(x)}g_{{\bf k}}(x)dx. 
\end{eqnarray}
Using the  boundary condition $\tau(\pm L) = 0$, we should have $D\int_0^L {\partial \over \partial x}\tau(x)dx = -D\tau(0)$; see Eq. \ref{eq-Dtau0}. 
We solve this problem using the Jacobi-Anger expansion again and obtain (with $I_n(-A) = (-1)^n I_n(A)$)
\begin{eqnarray}
	-D\tau(0) &=& DB_0I_0(U_a)I_0(U_b) L + DB_0\sum_{\bf k^{\prime}\ne 0} I_{\bf k^{\prime}} g_{\bf k^{\prime}}(L) \nonumber \\ 
	&-&I_0^2(U_a)I_0^2(U_b) {L^2 \over 2} - I_0(U_a)I_0(U_b) \sum_{{\bf k}^{\prime} \ne 0} k_{n^{\prime},m^{\prime}}(L)\nonumber\\
	&-& \sum_{\bf k\neq 0,k^{\prime}} I_{\bf k}I_{\bf k^{\prime}} \int_0^L e^{i{\bf k^{\prime}} \cdot {\bf r} x}g_{{\bf k}}(x)dx,
\end{eqnarray} 
where
\begin{eqnarray}
	k_{n,m}(L) 
	&=& I_n(-U_a)I_m(-U_b) \int_{0}^L \exp(i{\bf k} \cdot {\bf r} x) x dx \nonumber \\ 
	&=& I_n(-U_a)I_m(-U_b)(L + {g_{\bf k}(L) \over (i{\bf k} \cdot {\bf r})^2}). 
\end{eqnarray}
Using $|g_{\bf k}(L)| \le 2/|n/a+m/b|$, which is a bounded function when $n,m \ne 0$, we expect that in the large $L$ limit, $k_{n,m}(L) \rightarrow I_n(-U_a)I_m(-U_b)L$. 

In the same way we can calculate $\int_0^L e^{i{\bf k^{\prime}} \cdot {\bf r} x} g_{\bf k}(x)dx$, yielding that this term should, at most, proportional to $L$. This can be estimated using the following way
\begin{eqnarray}
	|\int_0^L e^{i{\bf k^{\prime}} \cdot {\bf r} x}g_{{\bf k}}(x)dx|\le\int_0^L |e^{i{\bf k^{\prime}} \cdot {\bf r} x}g_{{\bf k}}(x)|dx\le \frac{2L}{|{\bf k} \cdot {\bf r}|}.\nonumber\\
\end{eqnarray}
Collecting all these results, we  finally find that 
\begin{equation}
	-D\tau(0) = -I_0^2(U_a) I_0^2(U_b) {L^2 \over 2} + CL + \cdots, 
\end{equation}
where $C$ is related to $DB_0 I_0(U_a)I_0(U_b)$ and $-I_0(U_a) I_{n^{\prime}}(-U_a)I_0(U_b)I_{m^{\prime}}(-U_b)$ {\it etc}. The suspension points represent the next sub-leading terms with amplitude much smaller than the second term $CL$ in the large $L$ limit. In this way, in the large $L$ limit we have  $-D\tau(0) \rightarrow -I_0^2(U_a) I_0^2(U_b) {L^2 \over 2}$,  yielding
\begin{eqnarray}
	D^* &=& \lim_{L \rightarrow \infty} {L^2 \over 2\tau(0)} = {D \over \langle \exp(\Phi)\rangle \langle \exp(-\Phi)\rangle} \nonumber \\ 
    &=& {D \over I_0^2(U_a) I_0^2(U_b)}. 
    \label{eq-DeffJaJb}
    \end{eqnarray}
In this way we extend the LJ formula to the realm of quasi-periodic potentials, showing that it is still valid, as long as the diffusion constant is redefined properly using Eq. \ref{eq-Deff}, and the averaged is taken in a large enough interval using Eq. \ref{eq-average1}. Strikingly, we find that $D^*$ is always decreased with the increasing of incommensurate periods, since $|I_0(U)| > 1$ when $U \ne 0$.  

Several remarks are in order based on the above results.  Firstly,  the generalization of our results to much more complicated potentials are straightforward.  This result is a direct logical consequence that if we can approximate $U(x)$ using a periodic function, then the LJ formula should be applied immediately to study the diffusion constant $D^*$. In Ref. \cite{SangYang2025}, we show that for the following potential
\begin{equation}
U(x) = \sum_{i=1}^N U_i \cos({2\pi x \over a_i} + \theta_i),
\end{equation}
where $a_i$ are all pairwise incommensurate numbers and $\theta_i$ are their initial phase, then we should have 
\begin{eqnarray}
	D^* = {D \over \prod_{i=1}^N I_0^2(U_i)}, 
    \label{eq-DeffNUi}
\end{eqnarray}
which is independent of the phases $\theta_i$. We see that the cross terms from the interference between different periods are not presented strictly anymore. 
Secondly,  for all these general potentials we should always have 
\begin{equation}
	-\tau(x) = -{L^2 -x^2 \over 2D^*} + \text{sub-leading terms}. 
\end{equation}
With this result we find that when $L$ and $x$ are large enough (assuming $|x| < L$)
\begin{equation}
\lim_{x\rightarrow \infty} {\partial^2 \tau(x) \over \partial x^2}  =-{1\over D^*},
\end{equation} 
following Eq. \ref{eq-DUx0}, which is another way to define $D^*$. The sub-leading terms will become insignificant when $x$ approaches infinity.  From Eq. \ref{eq-St0}, it is also expected that $D^*$ may also be reflected from the long time tail of the survival probability. Thirdly, our result means that we can generalize the results to much more general potentials, in which the only requirement is the existence of the following limit  
\begin{equation}
\lim_{L \rightarrow \infty} {1\over L} \int_0^L \exp(\pm \Phi(x))dx = \text{finite}, 
\end{equation}
for which reason the results should have broad applicability in Brownian motion. Obviously, when $\Phi(x)  = kx^2$ for a trapped harmonic potential,  the above limit is either zero or infinity, thus $D^* = 0$. Finally, with the results in Ref. \cite{SangYang2025}, we can obtain $p(x, t|x_0)$ by solving the Fokker-Planck equation analytically, which can be used to calculate the survival probability $s(t, x_0)$, distribution of first passage time $\rho(t, x_0)$ and the associated MFPT, following Eq. \ref{eq-stx0} and Eq. \ref{eq-taux0}, with the aid of Jacobi-Anger expansion. This may lead to intriguing distribution of the first passage time $\rho(t, x_0)$ and its higher-order moments,  which deserve to be explored in future.  

The significance of the above results is that, we prove exactly that the LJ formula, if properly redefined, can be used to study the diffusion in much more general potentials. This can greatly broaden the applicability of this formula to study the Brownian motion in more general potentials. As compared with Ref. \cite{SangYang2025}, we prove  this conclusion exactly in terms of MFPT. We hope this result and the method to exactly calculate $\tau(x)$  can find immediate applications in Brownian motion in some concrete physical systems with some general potentials, in which the strict requirement of periodicity is abandoned, thus the cost of experimental realization is greatly reduced.  Intriguing applications of our theory may also include the anomalous diffusion in periodic potentials and quasi-periodic potentials \cite{Gitterman2000Mean, Dybiec2006Levy}. This  formula can also be used to study the giant diffusion in Brownian motion with tilted potentials \cite{Reimann2001giant, Reimann2008Weak, Reimann2002diffusion, Iida2025Universality}. 

\textit{Acknowledgments}:  We thank Sang Yang and Dr. Wencheng Ji for valuable discussion. This work is supported by the Strategic Priority Research Program of the Chinese Academy of Sciences (Grant No. XDB0500000),  and the Innovation Program for Quantum Science and Technology (2021ZD0301200, 2021ZD0301500). 


%

\end{document}